\documentstyle[sprocl]{article}
\input epsf
\newcommand{\Realno}{I\!\!R}%defines symbol for real numbers
\newcommand{\Integerno}{Z\!\!\!\!Z}%defines symbol for integer numbers
\def\openone{\leavevmode\hbox{\small1\kern-3.3pt\normalsize1}}
\def\lesssim{\mathrel{\mathpalette\vereq<}}
\def\vereq#1#2{\lower3pt\vbox{\baselineskip1.5pt \lineskip1.5pt
\ialign{$\hfill##\hfil$\crcr#2\crcr\sim\crcr}}}
\def\gtrsim{\mathrel{\mathpalette\vereq>}}
\begin{document}
 \title{THE THEORY OF %
BOUNDARY CRITICAL PHENOMENA$^\dagger$}%
\footnote[0]{$^\dagger$This article is based on an invited talk presented
at the {\em Third International Conference ``Renormalization
Group --- 96''\/}, Dubna, Russia, August 26 -- September 1,
1996.}

 \author{ H.~W.~DIEHL }
 \address{Fachbereich Physik, Universit\"at-Gesamthochschule Essen,\\%
D-45117 Essen, Federal Republic of Germany}
 \maketitle
\abstracts{%
An introduction into the theory of
boundary critical phenomena and the application of
the field-theoretical renormalization group method
to these is given.
The emphasis is on a discussion
of surface critical behavior
at bulk critical points
of magnets, binary alloys, and
fluids. Yet a multitude of related
phenomena are mentioned.
The most important distinct surface universality
classes that may occur for a given universality
class of bulk critical behavior are described, and
the respective boundary conditions
of the associated field theories are discussed.
The short-distance singularities of
the order-parameter profile in the diverse asymptotic regimes
are surveyed.
}
 \section{Introduction}

Until the end of the 70ies at least the subject of boundary
critical phenomena\cite{Bin83,Die86a}
attracted only rather limited attention,
even though a number of pioneering papers%
\footnote{Due to lack of space and the great number of papers %
concerned, it will not be possible to
present here an adequate account
of these earlier contributions and their importance.
For the same reasons I shall not in general
be able to refer to many original papers. My citation philosophy
will be to refer, wherever possible, to appropriate review articles
(such as \cite{Bin83,Die86a})
by whose extensive lists of references the interested reader can easily
trace the literature. I apologize to all colleagues whose papers could
not explicitly be cited here.}
had already been
written on it. This was due to several
reasons. Perhaps the most important one was the complete lack of
sufficiently precise experimental work. Such work even
seemed beyond reach of the then available experimental
possibilities. Another reason was that the powerful machinery of
the field-theoretical renormalization group (RG) approach had not
yet been extended to systems with boundaries, and hence could
be utilized for systematic studies of
boundary critical
phenomena\cite{Die86a}$^{\mbox{\scriptsize --\,}}$\cite{DN86}
only later.
Finally, also the seminal work of Belavin, Polyakov,
and Zamolodchikov\cite{BPZ84}
on two-dimensional conformal field theories
had yet to be performed and extended to
systems with surfaces.\cite{Car84}

Today the situation has changed considerably.
The subject has become a very active research field,
attracting many scientists with
a broad range of diverse backgrounds.
This includes the experimental and theoretical
condensed matter physicist who is interested in the critical
behavior of magnets and alloys with free surfaces%
\cite{Bin83,Die86a,Dos92}
just as well as the
physical chemist investigating the critical adsorption
of binary fluids on walls and interfaces,\cite{Law94} or the
theoretical high-energy physicist studying field
theories on manifolds with boundaries.\cite{MO92} Further, there
is a wealth of related phenomena. The reader can get a first
impression from the following incomplete list of topics
belonging to the realm of boundary critical
phenomena:
\begin{list}{(\roman{enumi})}{\usecounter{enumi}}
\item critical behavior of magnets and alloys with free
surfaces,\cite{Bin83,Die86a,Dos92}
\item critical adsorption of fluids on walls and
interfaces,\cite{Law94}$^{\mbox{\scriptsize --\,}}$\cite{FD95}
\item critical adsorption of polymers on walls,\cite{Eis93}
\item percolation transitions in systems with boundaries,\cite{DL89}
\item quantum impurity problems
(such as the Kondo effect),\cite{FLS95}
\item Schr\"odinger representation in
renormalizable field theories,\cite{Sym81}
\item Casimir effect,\cite{Kre94}
\item relaxation processes starting from initial
non-equilibrium states,\cite{JSS89,Jan92}
\item dynamical critical phenomena in systems with
boundaries.%
\cite{DD83b,DD85}$^{\mbox{\scriptsize --\,}}$\cite{WD95}
\end{list}

The common feature of these phenomena is that their physics
on large length and time scales is described by nearly critical
(or `massless') field theories on manifolds with boundaries.
These boundaries may be genuine surfaces
[as in examples (i), (iv), (vii), (ix)], walls
[as in (ii), (iii), (vii)],
or interfaces separating a nearly critical phase
(such as a binary mixed fluid near its consolute point)
from a non-critical spectator phase (such as vapor), where
the latter example
applies to the case of critical adsorption (ii)
of a binary fluid mixture
at its critical end point.%
\cite{Wid77}$^{\mbox{\scriptsize --\,}}$\cite{FU90b}
In
quantum impurity problems\cite{FLS95} [topic (v)], the role of
the boundaries is played by the impurities, which turn out to be
equivalent to boundary conditions of the resulting
effective 1+1 dimensional field theories.
In the examples (vi) and (viii) one is dealing with the time evolution
from a given initial state at time $t_i$; here
the space-time hyperplane $t=t_i$ is the analog of a
boundary.\cite{Sym81,Jan92}

The purpose of this paper is to give a brief introduction
into the field of boundary critical phenomena and
to survey some recent pertinent RG results.
The emphasis will be laid on a discussion of topics
(i) and (ii). This also serves to explain the basic questions
arising in, as well as general aspects of, boundary critical phenomena.
In addition, typical theoretical strategies will be described
and some representative experimental results mentioned.
A detailed exposition of the other
topics, (iii)--(ix), is beyond the scope of the present article.
The reader should consult the cited literature for more
information about  these.

\section{Surface effects}

Owing to their conceptual simplicity,
ferromagnets are a convenient starting point.
In studies of bulk critical behavior,
two important properties of real ferromagnets are usually ignored:
their {\em finite size\/} and the {\em presence of surfaces\/}.
To gain insight into the significance of these properties,
consider a ferromagnet of finite linear
extension $L$, where $L$ is large in comparison
with the characteristic
microscopic scale,  the lattice constant $a$.
 For concreteness, we take a $d$ dimensional cube.
Its $2d$ faces make up the boundary of total surface area $A=2dL^{d-1}$.
We presume that the microscopic interactions are of short range.
As a microscopic model one can imagine an Ising model
on the lattice $[0,L]^d\cap\Integerno^d$, with a Boltzmann
factor $\exp(-{\cal H\/}_{\mbox{\scriptsize lat}})$ of the form
\begin{equation} \label{latHam}
{\cal H\/}_{\mbox{\scriptsize lat}}
=\frac{1}{2}\sum_{{\mbox{\boldmath$i$}}\ne
{\mbox{\boldmath$j$}}}K_{{\mbox{\boldmath$i$}}{\mbox{\boldmath$j$}}}\,
s_{{\mbox{\boldmath$i$}}}\, s_{{\mbox{\boldmath$j$}}}\;.
\end{equation}
Here ${\mbox{\boldmath$i$}}$ labels the sites, and
$s_{{\mbox{\boldmath$i$}}}=\pm 1$ are spin variables.
Away from the surface, the interaction constants are translation
invariant,  $K_{{\mbox{\boldmath$i$}}{\mbox{\boldmath$j$}}}=
K({\mbox{\boldmath$i$}}\!-\! {\mbox{\boldmath$j$}})$,
but close to it $K_{{\mbox{\boldmath$i$}}{\mbox{\boldmath$j$}}}$
will in general also depend on the distance of
the bond's center from the surface.
Restricting ourselves to nearest-neighbor (nn) interactions, we assume
that $K_{{\mbox{\boldmath$i$}}{\mbox{\boldmath$j$}}}$
takes the value $K_1$ for all
nn bonds between two surface sites, the `bulk value' $K$ for all other
nn bonds, and is zero otherwise. It is understood that both $K$ and $K_1$
are ferromagnetic $(<0)$. No restriction is imposed
on the boundary variables $s_{{\mbox{\boldmath$i$}}}$
(free boundary conditions).\footnote{If required,
pure bulk phases with up or down magnetization are selected
by means of a homogeneous magnetic field $H$ which approaches
$\pm 0$ after the thermodynamic limit has been taken.}

The presence of surfaces entails that
thermal averages of local densities deviate
from their values deep inside the sample. Far away
from the critical point $K=K_c$, this disturbance
can penetrate into the sample only up to a distance
comparable with the interaction range, for
the correlation
length $\xi$ is of the same (microscopic) order.
However, as the temperature $T$ approaches the bulk critical temperature
$T_c$, $\xi$ grows, getting macroscopically large. Hence the boundary region
affected by the surface acquires a thickness of the same macroscopic order.
An obvious first consequence is:
\begin{itemize}
\item
{\em Local densities such as the order parameter
$m({\mbox{\boldmath$x$}})$
or the energy density $\varepsilon({\mbox{\boldmath$x$}})$
become inhomogeneous on the scale of $\xi$.}
\end{itemize}
In our case, $m({\mbox{\boldmath$x$}})$
and $\varepsilon({\mbox{\boldmath$x$}})$ correspond
to (coarse-grained versions of) the local
magnetization $m({\mbox{\boldmath$i$}})=
\langle s_{{\mbox{\boldmath$i$}}}\rangle$ and
energy
\begin{equation}
\varepsilon({\bf i}) =
\sum_{\mbox{\footnotesize nn bonds of }{\mbox{\boldmath$i$}}}\langle
s_{{\mbox{\boldmath$i$}}}s_{{\mbox{\boldmath$j$}}}\rangle\;,
\end{equation}
respectively.

The position-dependence of the magnetization
for $T<T_c$ along a line perpendicular to one pair of faces,
and sufficiently far away from the others,
is shown schematically in Fig.~1 for the case
$\xi\ll L$. 
\begin{figure}[h]
\def\epsfsize#1#2{0.6#1}
\centerline{%
\epsfbox{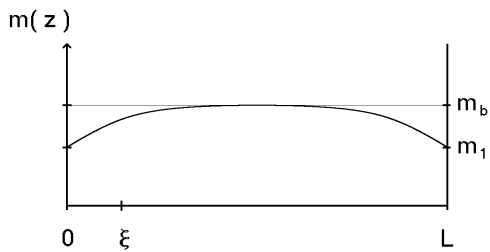}%
}
\caption{Schematic order-parameter profile $m(z)$ between
two parallel faces, where $z$ is the distance from the left face.
Up to exponentially small correction terms
due to the faces not shown, the profile
is independent of the coordinates perpendicular to $z$.}

%\medskip
\end{figure}
Deep inside the bulk,
$m({\mbox{\boldmath$x$}})$ agrees with the bulk value $m_b$ up to
corrections $\sim \exp(-z/\xi)$ (and analogous ones
with $z$ replaced by the distance to the other faces).
Owing to the reduced coordination number at
the surface, one expects
the profile to {\em decrease\/} upon approaching the
surface provided the
interactions are not too strongly enhanced near
the surface.
However, it is also possible that the magnetization
profile {\em bends upwards\/}
near the surface. This may happen when the enhancement of surface
interactions is sufficiently strong, or when the magnetization 
is locally increased by a magnetic
field acting only on spins in the vicinity of the surface.
As we shall see later,
there may even be changeovers between different
asymptotic forms of large-scale behavior
in the region $z\lesssim \xi$ if certain surface-related
crossover lengths become much larger
than the lattice constant $a$. Of course,
on the scale of $a$, the behavior of the densities
depends on microscopic details (chosen interactions,
lattice type, etc). 

A second, equally obvious, consequence is 
\begin{itemize}
\item{\em the appearance of surface corrections in integrated densities.}
\end{itemize}
For example, the total order parameter per volume $V$ ($=L^d$)
can be written as
\begin{equation} \label{bulksurf}
\frac{1}{V}\int_{\cal M}\!d^dx\;
m({\mbox{\boldmath$x$}})\approx m_b+\frac{A}{V}\,m_s+\ldots 
\end{equation}
in the limit $L\to\infty$. Here ${\cal M\/}$ denotes the region $\subset \Realno^d$
occupied by the system.
The ellipsis stands for (edge) contributions $\sim L^{-2}$
and faster decaying ones. The quantity $m_s$  is called
surface {\em excess} magnetization. An analogous decomposition 
into a bulk term, $f_b$, and an excess contribution $f_s$, called
{\em surface free energy density}, holds for the reduced free energy
\begin{equation}
{F\over k_BT}\equiv -\ln\mbox{Tr\,} 
e^{-{\cal H\/}_{\mbox{\tiny lat}}}\;
\end{equation}
and other quantities.

Surface corrections like $AV^{-1}f_s$ are small for large $L$ as long as
$\xi\ll  L$. For temperatures $T$ sufficiently close to the pseudo-critical
temperature%
\cite{Fis71}$^{\mbox{\scriptsize --\,}}$\cite{Bin87}
$T_{L}=T_c+O(L^{-\nu})$ of the finite system,
the correlation length is of the same order as  $L$. Then
the separation $(\ref{bulksurf})$ into bulk and surface contributions
looses its meaning. Finite-size effects
become important.  These have been  surveyed  by
Dohm recently\cite{DohmRG96}.
Since our subject here is boundary effects, we will always
assume that  $\xi\ll L$. 
Hence we first take the thermodynamic limit $L\to\infty$ and
only then consider the approach to the critical point.
By this procedure finite-size effects are eliminated.
It amounts to the study of {\em semi-infinite\/} systems.
In most of the following we will restrict our attention to
these.

The critical behavior of bulk densities
is described by familiar power laws.
For the order parameter $m_b$, and the singular parts of
the energy density $\varepsilon_b$ and free energy $f_b$,
one has, in zero magnetic field,
\begin{equation}\label{bop}
m_b \approx M_-|\tau|^\beta\quad\mbox{as }\tau\to 0^-\,,
\end{equation}
and
\begin{equation}\label{bfe}
\varepsilon^{\mbox{\scriptsize (sing)}}_b\approx
{B_\pm\over 1-\alpha}\,|\tau|^{1-\alpha}\;,\quad
f^{\mbox{\scriptsize (sing)}}_b\approx
B_\pm\,|\tau|^{2-\alpha}\quad\mbox{as }\tau\to 0^\pm\,,
\end{equation}
respectively.  Here $\alpha$ and $\beta$ are two independent
standard critical exponents in terms of which all other
commonly introduced bulk critical exponents can be
expressed.\footnote{We assume that the bulk dimension
$d$ is between the upper and lower critical dimensions
$d^*$ and $d_*$, so that hyperscaling is valid.
Thus $\alpha=2-d\nu$ and $\beta=(\nu/2)(d-2+\eta)$,
where $\nu$ and $\eta$ are the usual correlation-length
and correlation exponents, respectively.}
$M_-$ and $B_\pm$ are nonuniversal (metric) factors.

It is natural to expect that local densities,
taken at points within a distance $z\ll\xi$ from the surface,
also display critical behavior characterized by
power-law singularities, as the bulk critical point is
approached. 
In other words, a third consequence should be
\begin{itemize}
\item{\em the appearance of new critical indices
characterizing the critical behavior of 
surface quantities.\/}
\end{itemize}
Thus,  for the local surface order parameter
 $m_1\equiv m(z\!=\!0)$, one anticipates that
\begin{equation} \label{m1}
m_1\sim| \tau | ^{\beta_1} \quad\mbox{as } \tau\to 0^-\,,
\end{equation}
where  $\beta_1$ in general should be different from $\beta$.

This is indeed the case. It suffices here to mention
just a few exemplary sources of evidence.
From exact work\cite{MW73} on the semi-infinite Ising model
in $d=2$ dimensions it is known that $\beta_1=1/2$,
which is to be compared with the Onsager value $\beta=1/8$.
Even mean-field theory\cite{LR75,Bin83,Die86a}
gives different values, namely $\beta_1=1$ and $\beta=1/2$,
independent of $d$. The first clear {\em experimental\/} evidence
for a value of $\beta_1\ne\beta$, to my knowledge,
is due to Alvarado et al.\cite{ACH82}
These authors investigated the temperature dependence of
$m_1$ at the (100) surface of a Ni ferromagnet
via spin-polarized low-energy electron
diffraction. They found
$\beta_1=0.825^{+0.025}_{-0.040}\,$, a value
much larger than the bulk exponent $\beta\approx 0.36$
of the three-dimensional $O(3)$ Heisenberg model,
with which the observed bulk critical behavior
agrees quite well. Their result for $\beta_1$
is in reasonable agreement with the theoretical estimate\cite{DN86}
$\beta_1(d\!=\!3,n\!=\!3)= 0.84\pm 0.01$,
which was obtained by combining the results of
an $\epsilon=4-d$ expansion to second
order\cite{DD80,RG80} with those of
a $d-2$ expansion\cite{DN86} to the same order. (Here $n$ stands
for the number of components of the order parameter.)

More recently, the method of X-ray scattering under grazing
incidence has been employed with impressive success
to study the surface critical
behavior of the binary alloy FeAl.\cite{MDPJ90,Dos92}
As suggested in the theoretical work of Dietrich
and Wagner,%
\cite{DW83}$^{\mbox{\scriptsize --\,}}$\cite{DH95}
this technique
lends itself  well to accurate investigations of
surface critical exponents. It enables one to
determine several surface critical exponents independently,
so that scaling relations can be checked.

FeAl has a rather rich phase diagram.\cite{Dos92}
The measurements of
Mail\"ander et al\cite{MDPJ90,Dos92}
were performed at the continuous bulk phase transition
between the high-temperature phase with
B2 ordering and the low-temperature phase with DO$_3$ ordering.
This transition involves a two-component order parameter;\cite{Bin91}
its bulk exponent $\beta$ should agree with the
value $\beta(3,2)\approx 0.345$ of the $O(2)$
$|\phi|^4$ model in $d=3$ dimensions.\footnote{The
Hamiltonian of an appropriate continuum
field theory for this bulk transition
clearly will not be totally $O(2)$ symmetric. However, symmetry-breaking
terms such as cubic anisotropies are believed to be irrelevant\cite{Aha76} for
$n=2$.} This appears to be the case, although the
precision of the experimental estimates is not sufficient to
discriminate between the Ising value\cite{ZJ89} $\beta_1(3,1)\approx 0.33$
and the one for $n=2$ given above. In any case,
a much larger value of $\beta_1$ was found again, namely\cite{MDPJ90}
$\beta_1=0.75\pm 0.02$. This result is close to the estimate\cite{Die86a}
$\beta_1(3,2)\approx 0.8$ obtained from the $\epsilon$
expansion.\cite{DD80,RG80}

Monte Carlo calculations for the Ising
case%
\cite{LB90}$^{\mbox{\scriptsize --\,}}$\cite{RC95}
also gave clear evidence
for values of $\beta_1\ne\beta$.

Summing up, it may be said that there is indisputable evidence
for the fact that the surface exponent $\beta_1$ generally is
different from the bulk exponent $\beta$. However, this is only
part of the story. A central element of the modern theory of
bulk critical phenomena is the division into
{\em (bulk) universality classes\/}.
An obvious question to be asked is whether a similar classification
can be accomplished for surface critical behavior at bulk critical
points. The answer is yes. We call such classes
{\em surface universality classes\/}. As it turns out,
\begin{itemize}
\item {\em for a given bulk universality class,
there exist in general
several distinct surface universality classes.\/}
\end{itemize}
To explain this we must consider specific models.
Instead of continuing to work with lattice models such as the one
introduced in Eq.~(\ref{latHam}), we will
now turn directly to appropriate
continuum models describing the large-scale physics.

\section{Surface universality classes}

We presume that on large scales no long-range
interactions must be taken into account.
This requires that the following
necessary conditions are fulfilled:
\begin{list}{(\Roman{enumi})}{\usecounter{enumi}}
\item
All pair interactions must be of short range
or at most have long-range parts that are irrelevant in
the RG sense (anywhere, i.e., both in the bulk and in the
vicinity of the surface).
\item All boundary-induced contributions to the interactions
must be of short range in the distance $z$ from the surface
or at most have parts of long range in $z$ that are irrelevant
in the RG sense. This must hold, in particular, for any one-body
potential associated with the boundary.
\end{list}
One-body potentials decaying as a power $z^{-\omega}$
as $z\to\infty$ occur
in the case of fluids bounded by walls. However, the
exponents $\omega$ usually are larger than the critical value
$\omega_c=\beta/\nu$ below which such
long-range tails must be expected to become
relevant (see, e.g., pages 210--213 of Ref.~\ref{Die86a}).

The large-scale physics of systems satisfying (I) and (II)
 (which the lattice model of the previous section evidently does)
may be expected to be described by a continuum field
theory with a local action of the form
\begin{equation}\label{gfHam}
{\cal H\/}=\int_{\cal M}\!dV\,{\cal L\/}({\mbox{\boldmath$x$}})
+\int_{\cal B}\!dA\, {\cal L\/}_1({\mbox{\boldmath$x$}})\;,
\end{equation}
where ${\cal L\/}({\mbox{\boldmath$x$}})$
and ${\cal L\/}_1({\mbox{\boldmath$x$}})$ are functions of
the order-parameter field
$\phi({\mbox{\boldmath$x$}} )$ and its derivatives. 
For a general $d$ dimensional manifold ${\cal M\/}$ with boundary
$\partial {\cal M\/}={\cal B\/}$, the representation-invariant volume
and area elements are given by\cite{MO92}
 $dV=\sqrt{g}\,d^dx$ and $dA=\sqrt{g_{\cal B}\/}\,d^{d-1}x_{\|}$,
respectively,
where $g(\bf x)$ is the determinant of the metric tensor while
$g_{\cal B\/}$ is its analog for the induced metric on ${\cal B\/}$.
We will restrict ourselves to manifolds ${\cal M\/}\subset\Realno^d$
in the sequel, and unless stated otherwise, we will also not consider
curved boundaries. Working with semi-infinite systems, our standard
choice of ${\cal M\/}$ will be the $d$ dimensional half-space
$\Realno^d_+\equiv \{({\mbox{\boldmath$x$}}_{\|},z)
\in \Realno^d\mid
0\le z<\infty\}$, with ${\cal B\/}$ given by the $z=0$ plane. Thus $dV$
and $dA$ can simply be read as $d^dx$ and $d^{d-1}x_{\|}$, respectively.

The `bulk density' ${\cal L\/}({\mbox{\boldmath$x$}})$
must be chosen in such a way that
a proper description of bulk critical behavior results. In the case
of a usual critical point of systems belonging to the universality
class of the isotropic $n$-vector model, this tells us to choose
\begin{equation}\label{Lbulk}
{\cal L\/}({\mbox{\boldmath$x$}})={\textstyle\frac{1}{2}}
\left[\nabla\phi({\mbox{\boldmath$x$}})\right]^2
+{\cal U\/}[\phi({\mbox{\boldmath$x$}})]
\end{equation}
with
\begin{equation}\label{Ubulk}
{\cal U\/}(\phi)={\textstyle\frac{1}{2}}\tau_0\phi^2+
{\textstyle\frac{1}{4!}}u_0|\phi|^4\;,
\end{equation}
up to symmetry-breaking terms.

In the overwhelming part of the following
it will not be necessary to include derivative terms in
the `boundary density' ${\cal L\/}_1$. Hence we choose
it to be of the form
\begin{equation}\label{L1}
{\cal L\/}_1({\mbox{\boldmath$x$}})
={\cal U\/}_1[\phi({\mbox{\boldmath$x$}})]\;.
\end{equation}
In fact, for the
Hamiltonians representing the surface universality
classes we will be concerned with, the choice
\begin{equation}\label{U1}
{\cal U\/}_1(\phi)={\textstyle\frac{1}{2}}
c_0\phi^2-h_{1,0}\,\phi
\end{equation}
is general enough. To understand why no $|\phi|^4$ boundary term
is included, one should note that the associated coupling constant
would have momentum dimension $\epsilon-1$. Thus
at least for sufficiently
small $\epsilon=4-d$, such a term should be irrelevant.%
\footnote{%
A $|\phi|^4$ boundary term is needed
to describe certain surface universality classes
of bulk {\em tricritical\/} systems\cite{DE87,ED88} in dimensions
$d\le 3$. Of course, in this case a $|\phi|^6$ bulk term
is required as well.}
A boundary contribution of the form $\phi\partial_n\phi$
cannot be excluded on the basis of power counting alone
since it involves a dimensionless coupling constant.
However, it can be ruled out on other
grounds
(it is redundant).\cite{DD81c,Die86a,DC91}

The phase diagram of the semi-infinite
$n$-vector model defined by
Eqs.~(\ref{gfHam})--(\ref{U1})
with $h_{1,0}=0$ is shown in Fig.~2.
\begin{figure}[h]
\def\epsfsize#1#2{0.58#1}
\medskip
\centerline{%
\epsfbox{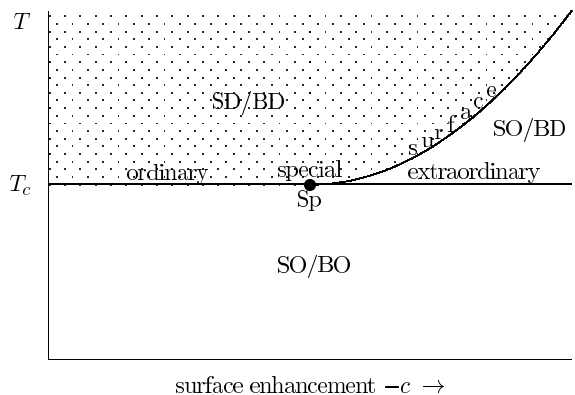}%
}
\caption{Phase diagram of the semi-infinite $n$-vector model
for $d-1>d_*(n)$.}
\medskip
\end{figure}
The vertical axis corresponds to $\tau_0$; in identifying
it with $T$, we have made use of the fact that $\tau_0$ is a linear
function of $T$ near $T_c$. The horizontal axis corresponds to $c_0$,
whose negative is a measure of the enhancement of the surface
interactions. The variable $c$ is proportional to
$c_0-c_{\mbox{\scriptsize sp}}$, where
$c_{\mbox{\scriptsize sp}}$ is chosen such that the point Sp
is located at $c=0$. The lines labeled
``ordinary'', ``extraordinary'', and ``surface'' represent
continuous phase transitions that have been given these
names.\cite{LR75} The lines meet at a multicritical point Sp,
which describes the ``special'' transition. The ordinary transition
occurs for {\em subcritical\/} surface enhancement $c>0$; it is a
transition from a surface-disordered (SD), bulk-disordered (BD) phase
to a surface-ordered (SO), bulk-disordered phase.
On lowering the temperature at fixed {\em supercritical\/}
enhancement $c<0$, one first enters a surface-ordered,
but bulk-disordered, phase at $T=T_{\mbox{\scriptsize s}}(c)$.
This so-called ``surface'' transition
is like a bulk transition of a $d-1$ dimensional system.
(The bulk correlation length of the $d$ dimensional
semi-infinite system remains {\em finite\/}
on the transition line, as long as $c$ stays
away from its critical value zero.)
As $T$ is lowered further at fixed $c<0$, the extraordinary
transition takes place.

At bulk criticality, $T=T_c$, we thus have {\em three distinct
types of surface transitions\/}: the ordinary, the special, and
the extraordinary one, provided
$d$ exceeds their respective lower critical dimensions (lcd).
The lcd of the surface, extraordinary, and special transitions
is $2$ for $n=1$, but $3$ for $n>1$.This is because a SO/BD phase
cannot appear (except for infinite surface enhancement),
unless the surface dimension $d-1$ is larger than
$d_*(n)=2-\delta_{n1}$. 
Hence, for the isotropic Heisenberg case $n=3$, only the
ordinary transition remains in three dimensions.\footnote{%
A SO/BD phase becomes possible in the Heisenberg case for $d>2$
when there is an easy-axis spin anisotropy at the surface
with a sufficiently enhanced surface interaction constant.
For a critical value of the enhancement
the transition temperature of the Ising-like surface transition
coincides with the $T_c$ of the Heisenberg bulk transition.
This defines an
anisotropic special point,\cite{DE84} which must not be confused with
the isotropic ones, Sp, we have in mind here.}
In the particular case $d-1=n=2$, a phase with quasi-long-range
order at the surface is possible.
This possibility will not be considered here. Note also that
the limit $n\to 0$ is known to describe the adsorption of
polymers on a wall.\cite{Eis93} For this problem an analog of
the multicritical point Sp, corresponding to an
adsorption threshold, exists
for $d\ge 2$.

For given $d$ and $n$, each one of the above-mentioned three
types of transitions of our model is
characteristic of a {\em separate\/}
surface universality class, also called ordinary, special, and
extraordinary, and described by
($h_{1,0}=0$) RG fixed points
\begin{equation}
{\cal P\/}^*_{\mbox{\scriptsize ord}}:\;
c^*_{\mbox{\scriptsize ord}}=\infty\;,\qquad
{\cal P\/}^*_{\mbox{\scriptsize sp}}:\;
c^*_{\mbox{\scriptsize sp}}=0\;,\qquad
{\cal P\/}^*_{\mbox{\scriptsize ex}}:\;
c^*_{\mbox{\scriptsize ex}}=-\infty\;,
 \end{equation}
respectively, with the indicated fixed-point values of $c$.

Let us briefly summarize some important properties of these
transitions. 

\subsection{The ordinary transition}

At the ordinary transition there is a {\em single relevant surface
scaling field\/} $g_1$, which varies $\sim h_{1,0}$ for small $h_{1,0}$.
All conventionally defined surface critical exponents can be expressed
in terms of its scaling index
$\Delta_1^{\mbox{\scriptsize ord}}/\nu$
and two independent bulk exponents. For example, the critical
surface pair correlation function, the singular part of the surface
energy density, and $m_1$ behave as
\begin{equation}
\left\langle\phi({\mbox{\boldmath$x$}}_{\|},0)\,
\phi({\bf 0},0)\right\rangle_{T=T_c}
\sim |{\mbox{\boldmath$x$}}_{\|}|^{-(d-2+\eta_{\|})}\;,
\end{equation}
\begin{equation}
\label{varepsb}
\varepsilon_1^{\mbox{\scriptsize (sing)}}
\approx \hat{B}_{\pm} |\tau|^{2-\alpha}\;,
\end{equation}
and
\begin{equation}\label{scfm1}
m_1\approx|\tau|^{\beta_1^{\mbox{\tiny ord}}}\,\sigma_\pm
(g_1|\tau|^{-\Delta_1^{\mbox{\tiny ord}}})\;,
\end{equation}
with
\begin{equation}\label{scallaw}
\beta_1=\frac{\nu}{2}\,(d-2+\eta_{\|})=(d-1)\nu-\Delta_1\;.
\end{equation}

The exponent values of $\beta_1$ given in the previous section
all referred to the ordinary transition. The agreement of the field-theory
es\-ti\-mates\cite{Die86a,DD80,DD81c,DN86,RG80,DS94} with those obtained
by other means
is quite good, though one is far away from the precision achieved in the bulk
case\cite{ZJ89} because the field-theoretical calculations have been carried
out only to two-loop order so far. For the Ising case, one finds
$\beta_1^{\mbox{\scriptsize ord}}(3,1)\approx 0.8$.
Comparisons with
recent Monte Carlo and other estimates may be
found in Refs.~\ref{LB90}, \ref{RW95}, and \ref{DS94}.
Experiments with binary liquid mixtures yielded the estimate\cite{SF86}
$\beta_1^{\mbox{\scriptsize ord}}=0.83\pm 0.05$. (For another
recent experiment, see Ref.~\ref{BPH93}.)

Since scaling laws such as (\ref{scallaw}) can be derived quite generally
from the RG equations\cite{DD80,Die86a}, they must hold
to any order of the $\epsilon$ expansion (and the
$d-2$ expansion\cite{DN86}, for $n> 2$). From the RG equations
one can also conclude that scaling functions such as $\sigma_\pm$
are universal up to a fixing of scales. In the bulk case there are
two nonuniversal metric factors to be fixed (``two-scale-factor universality'').
Here there is one additional
nonuniversal factor,
associated with the relevant surface scaling field $g_1$.
Hence we have a {\em  $(2+1)$-scale-factor universality\/}.\cite{DGS85}

To explain another important property in an elementary fashion, we note
that the  order parameter profile
takes the scaling form
\begin{equation}\label{OPprof}
m(z,\tau)\approx M_-|\tau|^\beta\,P(z/\xi) \quad\mbox{with}\quad
P(\zeta)\approx\cases{1&as $\zeta\to \infty$,\cr
p_0\,\zeta^{(\beta_1-\beta)/\nu}&as $\zeta\to 0$,\cr}
\end{equation}
where the short-distance singularity of $P(\zeta)$ follows from
the requirement that $m(z\ll\xi)$ has the
temperature dependence $\tau^{\beta_1}$ of $m_1$. Since
$\beta_1^{\mbox{\scriptsize ord}}>\beta$,
$P(0^+)$ vanishes. That is, {\em the order parameter satisfies an
asymptotic Dirichlet boundary condition\/}. This argument can be
generalized in a straightforward fashion to multi-point correlation
functions, both for $\tau >0$ as well as for $\tau <0$.
As first shown in Ref.~\ref{DD81c}, the short-distance
singularity $\phi_{z\to 0}\sim z^{(\beta_1-\beta)/\nu}$
can be systematically obtained by means of a short-distance
expansion in terms of boundary operators
(cf.\ Refs.~\ref{Die86a} and \ref{MO92}, and Sect.\ 4
below). Explicit results for the scaling function $P(\zeta)$ obtained
by RG-improved perturbation theory to one-loop order may
be found in Ref.\ \ref{DGS85}.

\subsection{The special transition}

This transition involves besides  the analog of $g_1$,
which we call $h_1$,
a {\em second relevant
surface scaling field\/}, $c$.
One is led to scaling forms such as
\begin{equation}
m_1(\tau,c)\approx |\tau|^{\beta_1^{\mbox{\tiny sp}}}
\psi_\pm (c|\tau|^{-\Phi},h_1|\tau|^{-\Delta^{\mbox{\tiny sp}}_1})
\;.
\end{equation}
Standard arguments based on crossover scaling
imply that the crossover exponent $\Phi$
governs the behavior of the line of surface transitions
near Sp, giving
$[T_{\mbox{\scriptsize s}}(c)-T_c]/T_c\sim |c|^{1/\Phi}$.

In mean-field theory $\beta_1^{\mbox{\scriptsize sp}}=
\Phi=1/2$, independent of $n$. The $\epsilon$ expansion is
known to second order. Setting $\epsilon =1$ in the corresponding
$O(\epsilon^2)$ expressions gives\cite{Die86a}
$\beta_1^{\mbox{\scriptsize sp}}(3,1)\approx 0.22\ldots 0.25$,
in reasonable agreement with various Monte Carlo
results,\cite{LB90}$^{\mbox{\scriptsize--\,}}$\cite{RC95}
which yielded values in the range
$0.19 \ldots 0.24$ (even though the error bars
reported in some of these works are much smaller).
The agreement is less satisfactory for $\Phi$.
Earlier simulations\cite{LB90} gave
$\Phi(3,1)\approx 0.59$, but more recent Monte Carlo
work\cite{RW95}  suggests a value $\lesssim 0.5$. The discrepancy with
the reported $\epsilon$ expansion estimate is presumably due to
the unusually large $O(\epsilon^2)$ term of $\Phi$.
This view is supported by the results
of a recent paper,\cite{DS94} in which
the massive RG approach for fixed dimension $d=3$ has been
extended to semi-infinite systems.
From a two-loop calculation and subsequent
Pad\'e-Borel analysis the value  $\Phi(3,1)\approx 0.54$ was found.
If  such elaborate techniques are used to extrapolate the
$\epsilon$ expansion to $d=3$, one can get down to similarly small
estimates of $\Phi$. To significantly improve the
accuracy of such field-theory estimates,
calculations to higher loop orders are required.

The scaling form (\ref{OPprof}) of the order-parameter profile carries
over to the present $c=0$ case (with  $n=1$), but now the exponent
$\beta_1^{\mbox{\scriptsize sp}}-\beta$ of the short-distance singularity
is negative for  $3\le d<4$. Hence the analog of $P(\zeta )$
tends to $+\infty$ as $\zeta\to 0$. Only when
$\beta_1^{\mbox{\scriptsize sp}}=\beta$ (as in mean-field theory and hence
for $d>4$), do we have a Neumann boundary condition $\partial_n\phi=0$.
Again,
these scaling considerations can be extended
to $N$-point correlation functions, and the short-distance singularities
can be derived in a systematic manner via a boundary-operator expansion
(cf.\ Sect.\ 4).

\subsection{Extraordinary and normal transitions and critical adsorption}

The special feature of the extraordinary transition is
{\em that there is spontaneous symmetry breaking and
long-range surface order both
below and above\/} $T_c$. Bray and Moore\cite{BM77} argued
that the important point is just that $m_1$ remains nonzero at the
transition. Instead of having long-range surface order caused
by spontaneous symmetry breaking for $T\gtrsim T_c$, one could
as well have surface order induced by a surface magnetic
field $h_{1,0}$. In other words, the surface critical behavior
of the extraordinary transition (supercritical enhancement
$c<0$ and $h_{1,0}=0$) should be
the same as in the case of subcritical enhancement $c>0$
with $h_{1,0}>0$. Applying scaling arguments to the latter case,
they predicted the behavior
$m_1^{\mbox{\scriptsize (sing)}}\sim |\tau|^{2-\alpha}$,
corresponding to $\beta_1^{\mbox{\scriptsize ex}}=2-\alpha$.

The case $c>0$, $h_{1,0}\ne 0$ is quite normal for
bounded fluids and binary fluid mixtures in contact with a wall,
since for these a surface ordering field (and other terms breaking the
$\Integerno_2$ symmetry) generically should be present
even at bulk coexistence. For this reason, the transition
with $h_{1,0}\ne 0$ and arbitrary $c<\infty$ has been termed
``normal''.\cite{Die94b,Die94a} Thus the claim is that the extraordinary
and normal transition are representative of one and the same surface
universality class, provided both are possible. (Note that
the l.c.d.\ of the normal transition is $d=2$, whereas it is $d=3$ for
the $n=1$ extraordinary transition.)\footnote{The reader should
be cautioned that the physically reasonable distinction between the
extraordinary and normal transitions is frequently not made in the
literature: It is a common, but unfortunate, practice to
use the name ``extraordinary'' for both the normal and extraordinary
transitions as well as for the associated universality class.}

In subsequent work by Burkhardt and the present author\cite{BD94}
it has been possible to demonstrate the extraordinary-normal
equivalence for the Ising case in an exact manner.
Likewise, the asymptotic behavior
at the normal and extraordinary transitions,
\begin{equation}\label{m1B}
m_1-m_1^{\mbox{\scriptsize (reg)}}\approx \tilde{B}_\pm
|\tau|^{2-\alpha}
\end{equation}
with
\begin{equation}\label{m1reg}
m_1^{\mbox{\scriptsize (reg)}}=m_{1,c}
+A_1\tau +A_2 \tau^2 +\ldots\;,
\end{equation}
is well established.\cite{BD94,DS93,Die94a} The ratios
of the coefficients $B_\pm$,
$\hat{B}_\pm$ and $\tilde{B}_\pm$ of the $|\tau|^{2-\alpha}$ singularities
in Eqs.\ (\ref{bfe}), (\ref{varepsb}), and (\ref{m1B}) are all given
by the same universal bulk ratio, i.e., $\tilde{B}_+/\tilde{B}_-=
\hat{B}_+/\hat{B}_-=B_+/B_-.$

In the perturbative RG approach
the normal-extraordinary equivalence
manifests itself as follows: If $h_{1,0}>0$, then
its renormalized analog, $h_1$, is driven to
$+\infty$ under the RG flow.
The variable $c$, on the other hand, tends to
$\pm\infty$ or stays at zero, depending on its initial value.
It turns out that the free propagator and the
zero-loop profile $m(z)$ for all these cases with $h_1=\infty$
are given by the same expressions and identical with their
analogs for $h_1=0$ and $c=-\infty$. Hence the
respective perturbation series agree to arbitrary order.
This shows that the limiting
probability distributions are the same; in this
sense there is just one fixed point for both
transitions.

The asymptotic behavior of the order-parameter profile
can be written in the scaling form (\ref{OPprof}),
but with different scaling functions
$P_\pm(\zeta)$ for $\tau\to 0^\pm$. These functions have
been computed by means of RG-improved perturbation theory
to one-loop order in $4-\epsilon$ dimensions.\cite{DS93}
The results, extrapolated to $d=3$, are in reasonable
agreement with Monte Carlo results.\cite{SDL94}
The short-distance behavior of the scaling functions,
\begin{equation}\label{Ppm}
P_\pm(\zeta)\begin{array}[t]{c}
\approx\\^{\zeta\to 0}\end{array}
\zeta^{-\beta/\nu}\left[ c_\pm
+a_{1,\pm}\, \zeta^{1/\nu}+a_{2,\pm}\,\zeta^{2/\nu}
+\tilde{b}_\pm\,\zeta^d+\ldots \right]\;,
\end{equation}
is in conformity with the $\tau$
dependence of $m_1$ given in Eq.~(\ref{m1B}).
The first three contributions correspond to the
terms regular in $\tau$ listed in Eq.~(\ref{m1reg});
the fourth one contains the $|\tau|^{2-\alpha}$ singularity.
This latter term may be understood as the contribution from
the component $T_{zz}$
of the stress energy tensor to the boundary
operator expansion\cite{Die86a,MO93}
(BOE)\footnote{The operators here should be read as renormalized
ones. In the sequel we will use the label `ren' to distinguish
renormalized from bare operators when necessary.}
\begin{equation}\label{BOE}
{\phi(\mbox{\boldmath$x$})\over
\langle\phi(\mbox{\boldmath$x$})\rangle_c}
\approx\sum_\lambda
C_\lambda (z)\,
{\cal O}_{\cal B}^\lambda(\mbox{\boldmath$x$}_{\cal B})
\end{equation}
of $\phi$ for $\mbox{\boldmath$x$}=
(\mbox{\boldmath$x$}_{\|},z)\to
\mbox{\boldmath$x$}^{\cal B}=
(\mbox{\boldmath$x$}_{\|},0)$.\cite{ES94}
Here $\langle\phi(\mbox{\boldmath$x$})\rangle_c
\propto z^{-\beta/\nu}$ is the
critical profile. The sum runs over a complete set
of boundary operators ${\cal O}_{\cal B}^\lambda$
with scaling exponents $\Delta_\lambda$.
The functions $C_\lambda(z)$ vary $\propto z^{\Delta_\lambda}$
at criticality. Consistency with Eq.~(\ref{Ppm}) requires
that
the boundary operator in (\ref{BOE}) with the
{\em smallest scaling dimension\/},
except the one operator $\openone$,
has $\Delta_\lambda =d$. This is $T_{zz}$.

To my knowledge, there is still no experimental system that
has been clearly demonstrated to have supercritical surface
enhancement. Thus it is fortunate that
extraordinary critical behavior can be seen at
normal transitions. A much studied phenomenon, which
occurs at such transitions, 
is the {\em critical adsorption of
fluids\/}.\cite{Law94,FdG78}$^{\mbox{\scriptsize --\,}}$\cite{FD95}
This occurs when, for example, a binary liquid mixture
 is brought to its bulk critical point in the presence of an external wall
(e.g., container wall) or other distinct physical interface.
For such a mixture, the order parameter is a composition variable.
The wall usually  favors one of the components, a property which translates
into the presence of a surface ordering field $h_{1,0}$.
Hence one gets back to our previous choice of Hamiltonian for
the normal transition. One important signature
of critical adsorption can
be read off from  Eq.~(\ref{Ppm}): the composition
varies as $z^{-\beta/\nu}$ for  $z\ll \xi$. A second one is
that the excess order parameter $m_s$ (corresponding to the
``total amount of adsorbed order'')
diverges as $\tau^{-(\nu-\beta)}$.
The analytical and Monte Carlo results given in Refs.~(\ref{DS93})
and (\ref{SDL94}) have proven useful for analyzing experiments
on critical adsorption. For example, values
for the universal amplitude $c_+$ in Eq.~(\ref{Ppm})
and for certain universal ratios involving integrals of $P_\pm$
have been extracted from experiments. These are in fair
agreement with the theoretical predictions\cite{Law94,FD95}
considering the still large error bars of both types of estimates.

\section{Field theory, boundary conditions, and short-distance singularities}

Since a detailed exposition of the field-theory approach to
surface critical behavior may be found Ref.~\ref{Die86a},
I will restrict myself here to giving a
brief summary of some essential points
for readers with a background in field theory.
As we have seen,
`operators' such as $\phi(\mbox{\boldmath$x$})$ generally have
different scaling dimensions for points {\boldmath$x$} in the interior,
$\mbox{int}\,{\cal M}\equiv{\cal M}\setminus {\cal B}$,
and on the boundary, ${\cal B.\/}$
For this reason we introduce separate bulk
and boundary source terms, defining
\begin{equation}
{\cal Z}[J,J_1]=e^{{\cal G}[J,J_1]}
\propto \int {\cal D}\phi\,\exp\!\left[-{\cal H}
+\int_{\cal M}\!dV\,J\phi +\int_{\cal B}\!dA\,
J_1\phi\right].
\end{equation}
The functional ${\cal G}$ generates the connected correlation
functions
$G^{(N,M)}=\langle\phi\ldots\phi\,
\phi_{\cal B}\ldots\phi_{\cal B}
\rangle^{\mbox{\scriptsize con}}$ of
$N$ fields $\phi(\mbox{\boldmath{$x$}}_i)$ with
$\mbox{\boldmath{$x$}}_i\in \mbox{int}\,{\cal M}$ and
$M$ fields $\phi_{\cal B}\equiv 
\phi(\mbox{\boldmath{$x$}}^{\cal B}_j)$ on the boundary.

Exploiting the invariance of ${\cal Z}$
under changes $\phi\to\phi+\delta\phi$ in a standard fashion gives
the `equations of motion'
\begin{equation}
J(\mbox{\boldmath$x$})=-\Delta \phi+{\cal U}'(\phi)\;,\quad
\mbox{\boldmath$x$}\in \mbox{int}\,{\cal M}\,,
\end{equation}
and
\begin{equation}\label{bcphi}
J_1(\mbox{\boldmath$x$}^{\cal B})=-\partial_n\phi+
{\cal U}_1'(\phi_{\cal B})
=-\partial_n\phi+c_0\phi_{\cal B}-h_{1,0}
\;,\quad\mbox{\boldmath{$x$}}^{\cal B}\in{\cal B}\,,
\end{equation}
where $\partial_n$ is the derivative along the inward normal.
The resulting boundary condition for the zero-loop profile
$m^{[0]}$
reads
$[\partial_z-c_0]\,m^{[0]}(z\!=\!0)=-h_{1,0}$;  the one for the
free propagator
$G=[-\Delta +U''(m^{[0]})]^{-1}
(\mbox{\boldmath$x$},\mbox{\boldmath$x$}')$
is of the Robin type,
\begin{equation}\label{bcG}
\partial_n G\big(\mbox{\boldmath$x$}^{\cal B},\mbox{\boldmath$x$}'\big)
={\cal U}_1''\big( m^{[0]}\big)\,
G\big(\mbox{\boldmath$x$}^{\cal B},\mbox{\boldmath$x$}'\big)
=c_0\,G(\mbox{\boldmath$x$}^{\cal B},\mbox{\boldmath$x$}')\;.
\end{equation}
It is well known that $G=G_b+G_s$, where
$G_b(|\mbox{\boldmath$x$}-\mbox{\boldmath$x$}'|)$ is its bulk
analog while 
$G_s\big(\mbox{\boldmath$x$},\mbox{\boldmath$x$}'\big)$ is an image
term. The former has an integrable singularity at
$\mbox{\boldmath$x$}=\mbox{\boldmath$x$}'$, giving rise
to the familiar ultraviolet (uv) bulk singularities in Feynman graphs.
The latter also becomes uv singular, but only for
$\mbox{\boldmath$x$}=\mbox{\boldmath$x$}'\in{\cal B}$. This
causes additional uv singularities in Feynman graphs. Owing
to the local form of the primitive divergencies,
they can be absorbed by {\em local
boundary count\-er\-terms\/}. The upshot is that
the required count\-er\-terms can be written
as a sum $\int_{\cal M}\mbox{CT}_b(\mbox{\boldmath$x$})
+\int_{\cal B}\mbox{CT}_1(\mbox{\boldmath$x$})$
of bulk and boundary contributions. Provided one
can convince oneself (e.g., by power counting)
that they all have the form of the interaction
terms included in the action, the theory is renormalizable.
In particular, our model defined by
Eqs.~(\ref{gfHam})--(\ref{U1}) turns out
to be (super-)renormalizable for $d=4$ ($d<4$).
One important difference with the
usual renormalization of infinite-space models
should be stressed, however: in general,
{\em one-particle {\em reducible\/}
renormalization parts may
occur\/}.\cite{Die86a,DD81b,DD83a,Sym81}

The counterterms needed to renormalize $G^{(N,M)}$
correspond to re\-pa\-ram\-e\-tri\-za\-tions of the form
\begin{equation}\label{bulkrp}
\phi=Z_\phi^{1/2}\phi^{\mbox{\scriptsize ren}}\;,\quad
\tau_0=\tau_c+\mu^2Z_\tau\tau\;,\quad
u_0=\mu^\epsilon Z_u u\;,
\end{equation}
\begin{equation}\label{surfrp}
\phi_{\cal B}=[Z_\phi Z_1]^{1/2}\,
(\phi_{\cal B})^{\mbox{\scriptsize ren}},\quad
c_0=c_{\mbox{\scriptsize sp}}+\mu Z_c c\;,\quad
h_{1,0}=\mu^{d/2} (Z_\phi Z_1)^{-1/2} h_1\;,
\end{equation}
where $\mu$ is the momentum scale.  
The $Z$ factors are meromorphic
in $\epsilon$, provided
dimensional regularization is used. In a theory
regulated by a momentum cut-off $\Lambda$,
one has
$c_{\mbox{\scriptsize sp}}\sim\Lambda$, analogous
to the familiar behavior $\tau_c\sim\Lambda^2$.
For the dimensionally regulated theory in fixed
dimensions $d<4$, a nonperturbative
shift of the form
$c_{\mbox{\scriptsize sp}}=
u_0^{1/\epsilon}\,{\cal C}(\epsilon)$
occurs,\cite{DS94} comparable to Symanzik's
mass-shift\cite{Sym73}
$\tau_c=u_0^{2/\epsilon}\,{\cal T}(\epsilon)$.

RG equations for the renormalized functions
$G^{(N,M)}_{\mbox{\scriptsize ren}}$
follow in a standard fashion and can be utilized
to derive their scaling forms and the scaling
relations for the exponents of
the special transition. The scaling functions
involve the scaling variables $c|\tau|^{-\Phi}$
and $h_1|\tau|^{-\Delta^{\mbox{\tiny sp}}_1}$.
In addition to the bulk correlation length $\xi$,
one has two surface-related
lengths $\xi_c\sim c^{-\nu/\Phi}$ and
$\xi_1\sim h_1^{-\nu/\Delta^{\mbox{\tiny sp}}_1}$
that become arbitrarily large as
the multicritical point Sp is approached.

For $c>0$ and $h_1=0$, a crossover to a behavior
characteristic of the {\em ordinary\/} transition occurs.
The information about the latter is contained
in the above-mentioned scaling functions.
However, it {\em cannot be deduced just from the
RG equations of the\/} $G^{(N,M)}_{\mbox{\scriptsize ren}}$.
To extract it, the limiting behavior of their
scaling functions for $c|\tau|^{-\Phi}\to \infty$
($\xi_c\ll \xi$) must be known. Setting $c$
to its value $c^*_{\mbox{\scriptsize ord}}=\infty$
at ${\cal P}^*_{\mbox{\scriptsize ord}}$ does not
really help because the bare and renormalized theories
satisfy the Dirichlet
boundary condition $\phi_{\cal B}^{\mbox{\scriptsize (ren)}}=0$
for $c_0=\infty$ and $c=\infty$ , respectively, so that
all functions $G^{(N,M)}_{\mbox{\scriptsize (ren)}}$
with $M>0$ vanish in this case. Hence, to determine
the critical behavior of these functions for $c>0$,
one must move
away from ${\cal P}^*_{\mbox{\scriptsize ord}}$,
considering large but finite values of $c$.
An alternative strategy is to infer
the thermal singularities of local properties
on ${\cal B}$ from those of their analogs
for small $z>0$, with $c=\infty$. This involves
knowledge about the corresponding
short-distance singularities.

Fortunately, a convenient way of obtaining the required
information has been found long ago.\cite{DD80,DD81c}
Using the boundary condition (\ref{bcphi}) [or (\ref{bcG}),
in perturbation theory], the operators
$\phi_{\cal B}$ in $G^{(N,M)}$ may be replaced
by $c_0^{-1}\partial_n\phi$. It follows that
the desired critical behavior is given by
the functions
$G^{(N,M)}_\infty \equiv
\langle\phi\ldots\phi\,
\partial_n\phi\ldots\partial_n\phi
\rangle^{\mbox{\scriptsize con}}$
with $c_0=\infty$.

The Dirichlet boundary condition of
the regularized theory
for $c_0=\infty$ holds
even if $h_{1,0}\ne 0$, provided
$h_{1,0}/c_0\to 0$ as $c_0\to\infty$.
In order to retain a nonvanishing relevant
surface field $g_1 \sim h_{1,0}$, the limit
$c_0\to\infty$ should be taken with
$h_{1,\infty} \equiv c_0^{-1}h_{1,0}$ fixed.
Noting that the term $\propto h_{1,0}$ in ${\cal H\/}$
can be rewritten as
$-h_{1,\infty}\int_{\cal B\/}\partial_n\phi$,
one sees that the relevant surface operator
to which $g_1\sim h_{1,\infty}$ couples
is $\partial_n\phi$.

Renormalization of the $c_0=\infty$ functions
$G^{(N,M)}_\infty$ can be achieved by means
of the reparametrizations\footnote{For $|c|<\infty$
one can easily see from the boundary condition
that a renormalized operator
$(\partial_n\phi)^{\mbox{\tiny ren}}$ can be defined by
$\partial_n\phi=Z_c[Z_\phi Z_1]^{1/2}
(\partial_n\phi)^{\mbox{\tiny ren}}$, in the dimensionally
regularized theory. However, for $c=\infty$ this is no longer
true.}
\begin{equation}
\partial_n\phi=[Z_\phi Z_{1,\infty}]^{1/2}\,
(\partial_n\phi)^{\mbox{\scriptsize ren}},\quad
h_{1,\infty}=\mu^{(d-2)/2} (Z_\phi Z_{1,\infty})^{-1/2}
h_{1,\infty}^{\mbox{\scriptsize ren}}
\end{equation}
and those given in Eq.~(\ref{bulkrp}).
The implied RG equations of the renormalized functions
$G^{(N,M)}_{\infty,\mbox{\scriptsize ren}}$
can be exploited in the usual manner to analyze the
critical behavior at the ordinary transition.\cite{Die86a,DD80,DD81c}
In particular, the scaling relations for the
critical exponents, and scaling forms such as (\ref{scfm1})
follow, with $h_{1,\infty}^{\mbox{\scriptsize}}$ playing the role of
$g_1$.

We close with a brief discussion of short-distance singularities (cf.\
pp.\ 190--202 of Ref.\ \ref{Die86a}).
Consider first the case of the {\em special\/} transition.
We are interested in the asymptotic $z$ dependence of the functions
$G^{(N,M)}_{\mbox{\scriptsize ren}}$ at criticality, $h_1=c=\tau =0$,
as one of the $N$ off-surface points, $\mbox{\boldmath$x$}=
(\mbox{\boldmath$x$}_{\|},z)$, approaches
$\mbox{\boldmath$x$}^{\cal B}=
(\mbox{\boldmath$x$}_{\|},0)$. The answer is provided by the BOE
\begin{equation}\label{BOEsp}
\phi^{\mbox{\scriptsize ren}}(\mbox{\boldmath$x$})
\begin{array}[t]{c}
\approx\\^{z\to 0}\end{array}
C(z)\,(\phi_{\cal B})^{\mbox{\scriptsize ren}}
(\mbox{\boldmath$x$}^{\cal B})\quad
\mbox{with }C(z)\propto z^{-(\beta-\beta_1^{\mbox{\tiny sp}})/\nu}\;.
\end{equation}
The form of $C(z)$ follows from its RG equation, implied by
those of $G^{(N,M)}_{\mbox{\scriptsize ren}}$. The exponent of $z$
can be rewritten as $(\beta_1-\beta)/\nu=\eta_\perp-\eta$, using
scaling laws.

The analog of this BOE for the functions
$G^{(N,M)}_{\infty,\mbox{\scriptsize ren}}$
at $h_{1,\infty}^{\mbox{\scriptsize ren}}=\tau =0$
reads\cite{DD81c}
\begin{equation}
\phi^{\mbox{\scriptsize ren}}(\mbox{\boldmath$x$})
\begin{array}[t]{c}
\approx\\^{z\to 0}\end{array}\label{BOEord}
C_\infty(z)\,(\partial_n\phi)^{\mbox{\scriptsize ren}}
(\mbox{\boldmath$x$}^{\cal B})\quad
\mbox{with }C_\infty (z)\propto 
z^{(\beta_1^{\mbox{\tiny ord}}-\beta)/\nu}\;.
\end{equation}
Obviously, this BOE applies equally well to
the $G^{(N,0)}_{\mbox{\scriptsize ren}}$ at $c=\infty$
and $h_1=\tau =0$. That the leading contribution now arises
from $\partial_n\phi$ rather than from $\phi_{\cal B}$
reflects the Dirichlet boundary condition of the renormalized
theory at $c=\infty$.

As one moves away from the fixed points
${\cal P}^*_{\mbox{\scriptsize sp}}$ and
${\cal P}^*_{\mbox{\scriptsize ord}}$, permitting
some relevant fields to be nonzero, the coefficient
functions appearing in the operator algebra,
in general, also depend on these scaling fields.
Further, additional contributions may occur.
Let us specifically consider what happens to
the short-distance singularities of the functions
$G^{(N,M)}_{\infty,\mbox{\scriptsize ren}}$
and the BOE (\ref{BOEord}) when a small
surface field $h_{1,\infty}^{\mbox{\scriptsize ren}}$
is turned on.
This problem was investigated
and solved by Symanzik\cite{Sym81} in 1981.
His results do not seem to have been recognized widely,
especially not in the solid state community, even
though they were reformulated in the
language of the BOE and generalized to manifolds
with curved boundaries in subsequent work.\cite{MO92}

Note, first, that the analog of Eq.~(\ref{bcphi})
for the $c_0=\infty$ functions
$G^{(N,M)}_{\infty,\mbox{\scriptsize ren}}$
reads
\begin{equation}\label{bcphiinf}
J_{1,\infty}({\mbox{\boldmath$x$}}^{\cal B})=
\phi_{\cal B}-h_{1,\infty}\;,
\end{equation}
where $J_{1,\infty}$ is the source associated
with $\partial_n\phi$. In the absence of $J_{1,\infty}$,
this reduces to the boundary condition
$\phi_{\cal B}=h_{1,\infty}$ for the regularized bare theory.
(Position-dependent boundary values $\phi_{\cal B}$ can be
imposed by retaining a nonzero $J_{1,\infty}$, as was actually
done by Symanzik.)
The naive analog of this boundary condition
for the renormalized theory,
$\lim_{\mbox{\boldmath$x$}\to
\mbox{\boldmath$x$}^{\cal B}}
\phi^{\mbox{\scriptsize ren}}
=h_{1,\infty}^{\mbox{\scriptsize ren}}$, does {\em not\/}
hold because of short-distance
singularities. Instead one has (setting $\mu =1$)
\begin{equation}
C_{\openone}(z)^{-1}\,
\phi^{\mbox{\scriptsize ren}}(\mbox{\boldmath$x$})
\begin{array}[t]{c}
\longrightarrow\\^{z\to 0}\end{array}
h_{1,\infty}^{\mbox{\scriptsize ren}}\;,
\end{equation}
where the choice $C_{\openone}(z)=
\int_{\cal B}dA\,
G^{(1,1)}_{\infty,\mbox{\scriptsize ren}}
(\mbox{\boldmath$x$},\mbox{\boldmath$0$})$ with
$h_{1,\infty}^{\mbox{\scriptsize ren}}=0$ can be made
(cf.\ pp.~16--17 of Ref.\ \ref{Sym81}).
The behavior of this function at
${\cal P}^*_{\mbox{\scriptsize ord}}$
is well
known\cite{DD80,DD81c,Die86a} and
follows directly from its
RG equation, explicitly given in
Ref.\ \ref{Sym81}. One finds
\begin{equation}\label{C1}
C_{\openone}(z)\propto z^{1-\eta_\perp^{\mbox{\tiny ord}}}\;.
\end{equation}
The exponent $1-\eta_\perp^{\mbox{\scriptsize ord}}=
(\Delta_1^{\mbox{\scriptsize ord}}-\beta)/\nu$ reflects the
different asymptotic scale dependencies
$\sim\ell^{-\Delta_1^{\mbox{\tiny ord}}/\nu}$
and $\sim\ell^{\beta/\nu}$ of
$h_{1,\infty}^{\mbox{\scriptsize ren}}$
and $\phi^{\mbox{\scriptsize ren}}$ in the infrared
limit $\ell\to 0$.

Thus the BOE becomes\cite{MO92}
\begin{equation}\label{sdsord}
\phi^{\mbox{\scriptsize ren}}(\mbox{\boldmath$x$})
\begin{array}[t]{c}
\approx\\^{z\to 0}\end{array}\label{BOEordh1}
C_{\openone}(z)\,h_{1,\infty}^{\mbox{\scriptsize ren}}
+ C_\infty(z)\,(\partial_n
\phi)^{\mbox{\scriptsize ren}}
(\mbox{\boldmath$x$}^{\cal B})+\ldots\;,
\end{equation}
where the ellipsis stands for less singular contributions
(arising from other boundary operators as well as from
the omitted additional dependence
of coefficient functions such
as $C_{\openone}$ and $C_\infty$ on the scaling variable
$h_{1,\infty}^{\mbox{\scriptsize ren}}\,
z^{\Delta_1^{\mbox{\tiny ord}}/\nu}$.)

The existence of the short-distance singularity (\ref{sdsord})
is a central issue of a recent Letter,\cite{RC96}
in which it is verified by means of Monte Carlo simulations
for the $d=3$ Ising case. In subsequent work
by the same authors the $d=2$ Ising case is studied.\cite{RC96}

The information acquired about short-distance singularities
can be utilized to predict the position dependence of
the order-parameter profile $m(z)\equiv\langle\phi\rangle$ in
various asymptotic regimes with $z\ll \xi$. At bulk criticality,
$\tau=0$, the scaling form of the profile, $m\equiv m_c$, reduces
to\cite{BL83,Die86a}
\begin{equation}\label{scfoppsp}
m(z)\approx z^{-\beta/\nu}\,F
\big(zh_1^{\nu/\Delta_1^{\mbox{\tiny sp}}},
ch_1^{-\Phi/\Delta_1^{\mbox{\tiny sp}}}\big)\;.
\end{equation}
The limits $F(\infty,z_2)$ and $F(z_1,0)$ of the scaling
function $F(z_1,z_2)$ must exist and be nonzero. Further, we know from
the BOE (\ref{BOEsp}) that $F(z_1,0)\sim
z_1^{\beta_1^{\mbox{\tiny sp}}/\nu}$
as $z_1\to 0$.\footnote{Note that
in Eq.~(3.227)
of Ref.\ \ref{Die86a}
the exponent of this power
contains an incorrect minus sign; at other places
of this reference the correct value is given.}
Consistency with the scaling form near
${\cal P}^*_{\mbox{\scriptsize ord}}$ requires that
\begin{equation}\label{Wfunc}
F\big(z_1\! =\!zh_1^{\nu/\Delta_1^{\mbox{\tiny sp}}},
z_2\!=\! ch_1^{-\Phi/\Delta_1^{\mbox{\tiny sp}}}\big)
\begin{array}[t]{c}
\sim\\^{z_2\to \infty}\end{array} 
W\big(zg_1^{\nu/\Delta_1^{\mbox{\tiny ord}}}\big)
\end{equation}
with\cite{Die86a}
\begin{equation}
g_1=h_1c^{-y}\;,\quad y=
{\Delta_1^{\mbox{\tiny sp}}-
\Delta_1^{\mbox{\tiny ord}}\over \Phi}=
{\gamma_{11}^{\mbox{\tiny sp}}-
\gamma_{11}^{\mbox{\tiny ord}}\over 2\Phi}\;,
\end{equation}
where we know from Eqs.~(\ref{C1}) and (\ref{BOEordh1}) that
$W(z_3)\sim z_3^{1-\eta_\perp^{\mbox{\tiny ord}}}$ as $z_3\to 0$.
In the scaling regime where Eq.~(\ref{scfoppsp})
holds (small $c$ and $h_1$),
both lengths $\xi_c\sim c^{-\nu/\Phi}$ and
$\xi_1\sim h_1^{-\nu/\Delta^{\mbox{\tiny sp}}_1}$
are large. Eq.~(\ref{Wfunc}) involves a
length $\xi_{\mbox{\scriptsize ord}}
\sim g_1^{-\nu/\Delta_1^{\mbox{\tiny ord}}}$,
which can be large or small.
The resulting power laws describing the
$z$ dependence of $m_c(z)$ in the various regimes
are
\begin{equation}
m_c(z)\sim\cases{z^{-(\beta-\beta^{\mbox{\tiny sp}}_1)/\nu}&
for  $z\ll \xi_c\;,\;\xi_1\;,\; \xi_{\mbox{\scriptsize ord}}$\,,\cr
z^{1-\eta_\perp^{\mbox{\tiny ord}}}&
for  $\xi_c\ll z\ll
\xi_{\mbox{\scriptsize ord}}$\,,\cr
z^{-\beta/\nu}&
for  $\xi_c\ll \xi_{\mbox{\scriptsize ord}}\ll z$\,.\cr}
\end{equation}

These findings are in conformity with unpublished one-loop results
for $m_c(z)$,
obtained some time ago by Ciach and myself.\cite{Alina}
Details can be found in a recent
paper by Ciach and Ritschel.\cite{Alina}

\section{Concluding remarks}

As I have tried to illustrate in this talk
by discussing prototypical examples of such phenomena,
the application of field-theoretical RG methods
to the study of boundary critical phenomena is well established.
It has led to deep insights, has provided a reliable
theoretical framework, and yielded predictions for
experimentally accessible quantities.
Research over the past two decades has revealed a
surprising wealth of such phenomena.

It is particularly encouraging to see that
the number of experiments devoted to careful
investigations of boundary critical phenomena
has been constantly increasing for some years.
Since more precise experimental data should be
available soon, an obvious challenge for
theorists is to improve the still rather moderate
accuracy of the field-theoretical predictions.
To this end, it would be very desirable to extend
the presently available two-loop results
for the $\epsilon$ expansion\cite{Die86a}
and the massive RG approach
in fixed dimensions\cite{DS94} $d<4$ to higher orders,
so that more precise field-theoretical estimates for
critical exponents and amplitude ratios can be
gained via Pad\'e-Borel resummation techniques.

While achieving greater accuracy is certainly one
important goal for future research, it
should also be emphasized
that only {\em relatively few\/} models for
boundary critical phenomena have been
thoroughly investigated so far.
Thus interesting new features may still have
to be discovered.

To illustrate this point, let me finish by
reporting a result established
recently.\cite{DLBD96,LD97}
As we have seen above, systems exhibiting boundary
critical behavior at bulk critical points also can
be divided into universality classes. These
surface universality classes depend on
the bulk universality class and
additional relevant surface properties.
Examples of the latter we have
come to know so far are: the short range
of the change of interactions caused by
the surface; their short-range nature;
whether the surface enhancement
is subcritical, critical, or supercritical;
and whether symmetry-breaking surface terms 
(with $h_{1,0}\ne 0$)
are present or not.

In the case of
antiferromagnets in a magnetic field (sufficiently
weak so that the bulk transition remains continuous)
and binary alloys with non-ideal stoichiometry
and a continuous order-disorder bulk transition,
it turns out that the surface universality class
in general also depends {\em on the orientation of
the surface plane with respect to the crystal axes\/}.
There exist both orientations that
preserve, as well as those that break, the symmetry with
respect to the
two types of sublattices.\cite{Sch93}
Depending on whether the orientation
is symmetry preserving or breaking,
ordinary and normal transitions
(corresponding to the ordinary and
extraordinary universality classes)
are expected to occur, respectively
(provided the enhancement is subcritical,
which seems reasonable to assume).\cite{DLBD96,LD97}
To appreciate this result one should note
that, on the level of a lattice description,
{\em no\/} ordering bulk and surface fields
are present, because these would correspond
to {\em staggered\/} magnetic fields, in antiferromagnetic
language. It should be possible to
check these predictions by means of experimental studies
of the continuous A2--B2 bulk transitions
in binary alloys such as FeCo\footnote{%
After completion of this article, new experimental results %
have been reported for FeCo.\cite{KDN+97} These experiments
gave clear evidence for the existence of an effective ordering surface
field. However, the crossover to normal surface critical behavior
could not be seen. Presumably, this requires a closer approach to $T_c$. } or FeAl. 

\section*{Acknowledgments}

I am grateful to my collaborators of the joint
papers mentioned in this talk, to Anja
Drewitz for producing Fig.~1, and to Reinhard Leidl for
a critical reading of the manuscript.
My work was partially supported by the Deutsche Forschungsgemeinschaft
via Sonderforschungsbereich 237 and the Leibniz program.

\section*{Note added in proof}

We mentioned above the result taken from Ref.\ \cite{Bin91} that the disorder-order transition of FeAl from the B2 to the DO$_3$ phase should belong to the universality class of
the O(2) $|\phi|^4$ model. Recently R.\ Leidl (U.\ Essen, unpublished)
has reanalyzed the problem within Landau theory. He found that the Landau theory
allows quadratic anisotropies. These are relevant and should drive the system to
the the $n=1$ (Ising) fixed point. Thus the transition should belong to the
universality class of the one-component $|\phi|^4$ model. Upon contacting Prof.\ Binder's group we learned that similar conclusions have been reached
in the paper by
W. Helbing, B. D\"unweg, K. Binder and D. P. Landau,
{\it Z.\ Physik\/}  {\bf  B 80}, 401 (1990), which corrects an earlier one by
B.\ D\"unweg and K.\ Binder, {\it Phys.\ Rev.\ }{\bf B 36}, 6935 (1987)
whose different conclusions are described in Ref.\ 48. We are grateful
to Prof.\ Binder and B.\ D\"unweg for kindly informing us about this matter.

\end{document}